%% file: main.tex
\documentclass[conference]{IEEEtran}
\IEEEoverridecommandlockouts

\usepackage{cite}
\usepackage{amsmath,amssymb,amsfonts}
\usepackage{algorithmic}
\usepackage{graphicx}
\usepackage{textcomp}
\usepackage{xcolor}
\usepackage{multirow}
\usepackage{float}
\usepackage{hyperref}
\usepackage{booktabs}
\usepackage{pifont}
\newcommand{\cmark}{\ding{51}}%

\usepackage{colortbl} 

\def\BibTeX{{\rm B\kern-.05em{\sc i\kern-.025em b}\kern-.08em
    T\kern-.1667em\lower.7ex\hbox{E}\kern-.125emX}}
\begin{document}

\title{Exploring Optimal Transport-Based Multi-Grained Alignments for Text-Molecule Retrieval\\
}

\author{\IEEEauthorblockN{Zijun Min\textsuperscript{1*}, Bingshuai Liu\textsuperscript{1*}, Liang Zhang\textsuperscript{1}, Jia Song\textsuperscript{1}, Jinsong Su\textsuperscript{1$\dagger$}, Song He\textsuperscript{2$\dagger$}, Xiaochen Bo\textsuperscript{2$\dagger$}}
\IEEEauthorblockA{
\textsuperscript{1}\textit{School of Informatics, Xiamen University, Xiamen, China} \\
\textsuperscript{2}\textit{Institute of Health Service and Transfusion Medicine, Beijing, China}\\
\{minzijun, bsliu, lzhang, songjia\}@stu.xmu.edu.cn, jssu@xmu.edu.cn, \{hes1224, boxiaoc\}@163.com}
\thanks{* Equal contribution.}
\thanks{$\dagger$ Corresponding author.}
}

\maketitle

\input{0-Abstract}


\input{1-Introduction}
\input{2-Related_Work}
\input{3-Method}

\input{4-Experiments}
\input{5-Conclusion}

\section{Acknowledgements}

The project was supported by the National Natural Science Foundation of China (Nos. 62276219 and 62472370) and the Public Technology Service Platform Project of Xiamen (No. 3502Z20231043).
We also thank the reviewers for their insightful comments.

\bibliographystyle{ieeetr}
\bibliography{ref}

\end{document}

%% file: 0-Abstract.tex
\begin{abstract}

The field of bioinformatics has seen significant progress, making the cross-modal text-molecule retrieval task increasingly vital. This task focuses on accurately retrieving molecule structures based on textual descriptions, by effectively aligning textual descriptions and molecules to assist researchers in identifying suitable molecular candidates. However, many existing approaches overlook the details inherent in molecule substructures. In this work, we introduce the \underline{O}ptimal T\underline{R}ansport-based \underline{M}ulti-grained \underline{A}lignments model (ORMA), a novel approach that facilitates multi-grained alignments between textual descriptions and molecules. Our model features a text encoder and a molecule encoder. The text encoder processes textual descriptions to generate both token-level and sentence-level representations, while molecules are modeled as hierarchical heterogeneous graphs, encompassing atom, motif, and molecule nodes to extract representations at these three levels. A key innovation in ORMA is the application of Optimal Transport (OT) to align tokens with motifs, creating multi-token representations that integrate multiple token alignments with their corresponding motifs. Additionally, we employ contrastive learning to refine cross-modal alignments at three distinct scales: token-atom, multitoken-motif, and sentence-molecule, ensuring that the similarities between correctly matched text-molecule pairs are maximized while those of unmatched pairs are minimized. To our knowledge, this is the first attempt to explore alignments at both the motif and multi-token levels. Experimental results on the ChEBI-20 and PCdes datasets demonstrate that ORMA significantly outperforms existing state-of-the-art (SOTA) models. Specifically, in text-molecule retrieval on ChEBI-20, our model achieves a Hits@1 score of 66.5\%, surpassing the SOTA model AMAN by 17.1\%. Similarly, in molecule-text retrieval, ORMA secures a Hits@1 score of 61.6\%, outperforming AMAN by 15.0\%.

\end{abstract}

%% file: 1-Introduction.tex
\section{Introduction}

\begin{figure}
    \centering
    \includegraphics[width=0.48\textwidth]{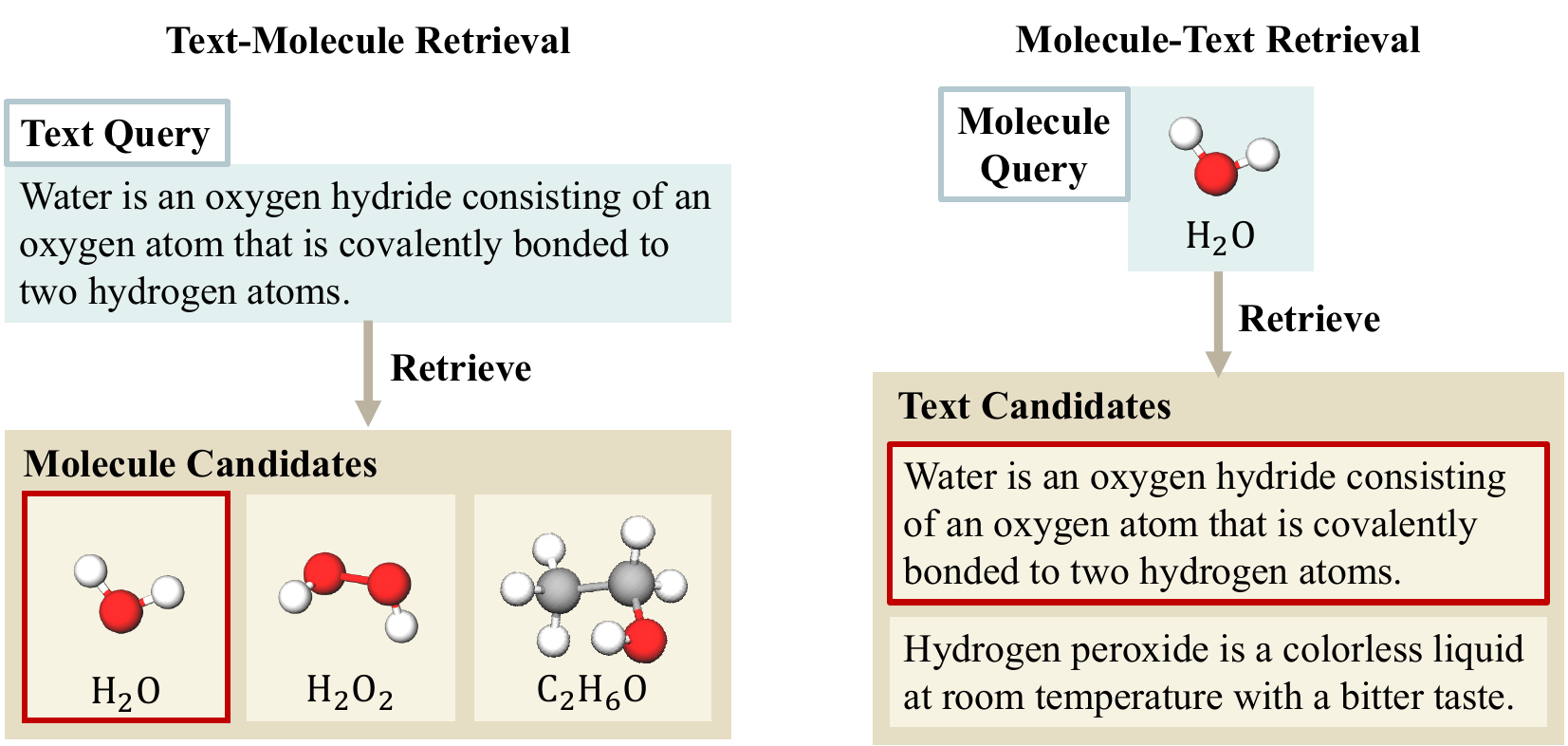}
    \caption{The text-molecule retrieval task is designed to retrieve molecules based on text queries, while molecule-text retrieval task does the opposite. The red box indicates the ground-truth retrieval result.}
    \label{fig:motivation}
\end{figure}

The rapid advancement of bioinformatics has led to the construction of numerous large-scale molecular databases, such as PubChem \cite{kim2016pubchem}. 
These databases play a crucial role in the discovery and synthesis of new drugs.
However, accurately retrieving desired molecules from these databases presents a significant challenge.
Therefore, aiming to retrieve molecules based on text queries, cross-modal text-molecule retrieval \cite{text2mol2021edwards} has become increasingly important.

Generally, existing studies mainly focus on the utilization of neural networks to learn representations of textual descriptions and molecules, and then calculating text-molecule similarities for retrieval. 
For example, several studies \cite{deep2022zeng, molt52022edwards, molxpt2023liu} resort to pretrained models based on Simplified Molecular Input Line Entry Specification (SMILES) \cite{smiles1998weininger} and text sequences. Alternatively, more studies like MoMu \cite{momu2022su} and MoleculeSTM \cite{moleculestm2023liu} represent molecules as 2D topological graphs, and then employ cross-modal contrastive learning to align molecular graphs and textual descriptions within a shared semantic space. 
Furthermore, AMAN \cite{adversarial2023zhao} utilizes adversarial learning to effectively bridge these two modalities, achieving state-of-the-art (SOTA) performance.

Despite these advancements, most studies overlook the detailed structural information essential for understanding molecular properties. 
Molecules are composed of atoms connected by chemical bonds, and their properties can be influenced by their structural motifs. 
In molecular chemistry, a motif is defined as a specific group of bonded atoms that follows a consistent and repeating pattern.
Ignoring such detailed structural information seriously limits the precision of retrieval results. The only exception is the recently proposed Atomas \cite{atomas2024zhang}, which applies clustering algorithms to extract representations at multiple granularities.

To address the above issue, we propose a novel text-molecule model with \textbf{O}ptimal T\textbf{R}ansport-based \textbf{M}ulti-grained \textbf{A}lignments \textbf{(ORMA)}.
Overall, our model contains a SciBERT-based text encoder and a GCN-based molecule encoder, to individually learn textual and molecular representations at multiple granularities for retrieval.

Concretely, we utilize the text encoder to process each input textual description, obtaining token and sentence representations. 
Simultaneously, we decompose each input molecule into motifs based on chemical rules, and then represent it as an undirected heterogeneous graph.
In this graph, each node represents either an atom, motif, or global molecule node, with two types of edges modeling the relationships between atom and motif, and motif and molecule, respectively.
Based on this graph, we employ the molecule encode to encoder the input molecule at multiple levels, obtaining atom, motif, and molecule representations.

Subsequently, we treat the tokens and motifs as independent distributions and employ optimal transport to achieve their alignments.
Furthermore, for each motif, we obtain a fused representation of its aligned tokens as the \textit{multi-token} representation.

With the previously mentioned cross-modal representations at different granularities, we introduce contrastive learning losses in our model training to achieve alignments at three levels: token-atom, multitoken-motif, and sentence-molecule. 
These losses aim to maximize similarity scores for matched text-molecule pairs while minimizing those for unmatched pairs, thereby enhancing alignments across different modalities. 
During inference of text-molecule retrieval, we calculate similarities at three levels, and combine them to retrieve the molecule with the highest similarity. 
In molecule-text retrieval task, we perform this process in the opposite direction.

To summarize, the main contributions of our work are three-fold:
\begin{itemize}
    \item We propose ORMA, designed to learn textual and molecular representations at multiple levels. It employs multiple losses to align cross-modal representations at different levels, thereby enhancing cross-modal alignments for retrieval.

    \item We model the alignments between tokens and motifs as an optimal transport problem to learn multi-token representations for each motif, facilitating subsequent multitoken-motif alignments. To the best of our knowledge, our work is the first to explore the multitoken-motif alignments in text-molecule retrieval.

    \item Experimental results on the ChEBI-20 and PCdes datasets demonstrate that our model achieves significant improvements over the state-of-the-art (SOTA) models. Specifically, for the text-molecule retrieval on ChEBI-20, the Hits@1 score of our model is 66.5\%, outperforming the SOTA model--AMAN by 17.1\%. Similarly, in the task of molecule-text retrieval, our model achieves a Hits@1 score of 61.6\%, surpassing AMAN by 15.0\%.
    
\end{itemize}

%% file: 2-Related_Work.tex
\section{Related Work}

To achieve high-quality text-molecule retrieval, most studies represent molecules as 1D sequences, 2D molecular graphs, or 3D molecular conformations. 
In the aspect of 1D molecule modeling, SMILES \cite{smiles1998weininger} has been widely used to represent molecular sequences, emerging many typical pre-training models, such as KV-PLM \cite{deep2022zeng}, MolT5 \cite{molt52022edwards}, and Text+Chem T5 \cite{unifying2023christofidellis}. 
Meanwhile, some studies switch their attention to 2D topological graphs, where atoms and chemical bonds are considered as nodes and edges, respectively. 
For example, studies like MoMu \cite{momu2022su} and MoleculeSTM \cite{moleculestm2023liu} employ cross-modal contrastive learning to align text and molecular graphs in a shared semantic space. Moreover, MolCA \cite{molca2023liu} introduces a cross-modal projector, while AMAN \cite{adversarial2023zhao} and \cite{2024arXiv241023715S} employ adversarial learning to bridge these two modalities effectively. 
As the extension of the above studies, several studies incorporate additional modalities to aid the alignments between text and molecules, such as MolFM \cite{molfm2023luo} incorporated knowledge graphs, and GIT-Mol \cite{git2024liu} incorporated images.
Recently, some studies focus on the spatial information in 3D molecular conformations. For instance, 3D-MoLM \cite{3dmollm2024li} adopts a similar architecture to MolCA \cite{molca2023liu}, using 3D conformations to represent molecules.

Despite these advancements, most methods only focus on global molecular information, often overlooking detailed information at the motif and atom levels. 
To integrate detailed information, multi-grained alignments have been explored in various natural language processing (NLP) \cite{fan2018multi, zhang2018alignment, zhang2017battrae, su2015bilingual, su2016convolution, zhang2016bilingual} and computer vision (CV) \cite{xclip2022ma, jin2023dicosa, wang2023ucofia} tasks, enabling models to effectively process complex information.
However, in the task of text-molecule retrieval, few studies pay 
attention to this technique. 
The only exception is Atomas \cite{atomas2024zhang} which applies a clustering algorithm to extract features at the atom, fragment, and molecule levels. 

Although with the same motivation as Atomas, we construct a hierarchical heterogeneous molecular graph to obtain atom, motif, and molecule representations. 
More importantly, we introduce optimal transport to fuse the representations of multiple tokens aligned with the same motif.
Through contrastive learning, we align text and molecules at token-atom, multitoken-motif, and sentence-molecule levels.
Subsequent experimental results show that our model outperforms Atomas, confirming the effectiveness of our model.

%% file: 3-Method.tex
\begin{figure*}
    \centering
    \includegraphics[width=1.0\textwidth]{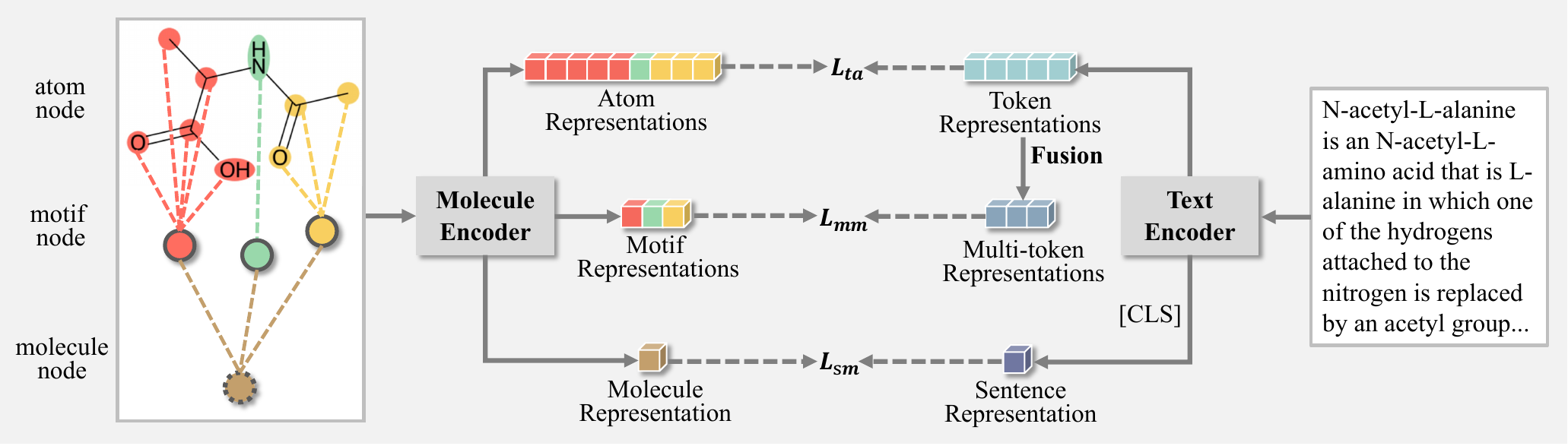}
    \caption{The illustration of our model. In short, our model comprises a text encoder and a molecule encoder and designs three alignment losses $\mathcal L_{ta}, \mathcal L_{mm}, \mathcal L_{sm}$ at token-atom, multitoken-motif, and sentence-molecule levels, respectively. On the left side of the figure is a molecular graph, where the nodes at the top represent atom nodes. In the middle, each node with the solid edge is a motif node, connected to the atom nodes it contains. At the bottom, the node with the dashed edge is the molecule node, connected to all motif nodes.}
    \label{fig:main_method}
\end{figure*}

\section{Our Model}
In this section, we will provide a detailed description of our model. 
We first present the main architecture of our model, which mainly contains a text encoder and a molecule encoder to learn representations at different granularities.
Next, we further introduce Optimal Transport (OT) to generate the fused representation of multiple tokens aligned with the same motifs.
To the best of our knowledge, our work is the first attempt to consider representations at the motif and multi-token levels.
Finally, we define a training objective that involves three alignment losses to achieve cross-modal multi-grained alignments.

\subsection{Main Architecture}
\label{sec:main_architecture}

As shown in Figure \ref{fig:main_method}, our model mainly consists of a molecule encoder and a text encoder. 
As implemented in previous studies \cite{text2mol2021edwards, adversarial2023zhao}, we also employ SciBERT \cite{scibert2019beltagy} to encode the input textual descriptions, obtaining representations at both sentence and token levels.
Compared with other pretraining models, SciBERT outperforms in encoding chemical textual descriptions, due to pretraining on a large-scale corpus of scientific publications.
Given an input textual description with $N_t$ tokens, we first concatenate a special [CLS] token at the beginning of the sequence.
Then, we input the sequence into the SciBERT encoder, where the learned $d$-dimensional representation of [CLS] serves as the sentence representation $h^s$, and those of remaining tokens are used as token representations: $\mathbf H^t=[h^t_1,h^t_2,\dots,h^t_{N_t}]\in\mathbb R^{N_t\times d}$.

To effectively encode the input molecule, we represent it as an undirected heterogeneous graph and use a GCN encoder to learn its representations at different granularities.
The molecular graph contains three types of nodes, constructed as follows:
(1) \textbf{Atom nodes}. 
Since each molecule consists of atoms connected by chemical bonds, 
we include each atom as an individual node, which enables us to fully capture the detailed structural information of the molecule.
(2) \textbf{Motif nodes}. 
In molecular chemistry, a motif is defined as a specific group of bonded atoms that follows a consistent and repeating pattern. 
Thus, we believe that motifs encode rich implicit semantic information, which is crucial for understanding the molecular properties described in the text.
Following previous study \cite{hierarchical2023zang}, we employ the BRICS algorithm \cite{brics2008degen} with an additional decomposition rule to extract motifs, all of which are also included as separate motif nodes.
(3) \textbf{Molecule node}.
We also include a global molecule node to facilitate the learning of a comprehensive molecule representation.
To effectively capture the relationships between nodes at different granularities, we consider two types of edges:
(1) \textbf{Motif-Atom Edges.} Each motif node is connected to its constituent atom nodes.
(2) \textbf{Molecule-Motif Edges.} The molecule node is connected to all motif nodes.

For example, in the molecular graph shown on the left part of Figure \ref{fig:main_method}, each node at the top represents an atom node, and each node with the solid edge in the middle is a motif node, connected to the atom nodes it contains.
Meanwhile, indicated as a node with the dashed edge, the molecule node is connected to all motif nodes.
This heterogeneous graph allows for enhancing the information sharing among atom, motif, and molecule nodes during the neighborhood aggregation of graph neural networks, thus facilitating multi-grained molecular representation learning.

Based on the above heterogeneous graph, we employ a 3-layer Graph Convolutional Network (GCN) ~\cite{gcn2016kipf} to learn node representations, which correspond to molecular representations at three granularities: 
(1) the atom representations $\mathbf H^a=[h^a_1, h^a_2, \dots, h^a_{N_a}]\in\mathbb R^{N_a\times d}$, where $N_a$ is denoted as the number of atoms,
(2) the motif representations $\mathbf H^m=[h^m_1, h^m_2, \dots, h^m_{N_m}]\in\mathbb R^{N_m\times d}$, with $N_m$ is the number of motif, and 
(3) the global molecule representation $h^g\in\mathbb R^{d}$.

\subsection{Optimal Transport-Based Multi-token Fused Representation Learning}
\label{sec:optimal_trans}

\begin{figure}
    \centering
    \includegraphics[width=0.48\textwidth]{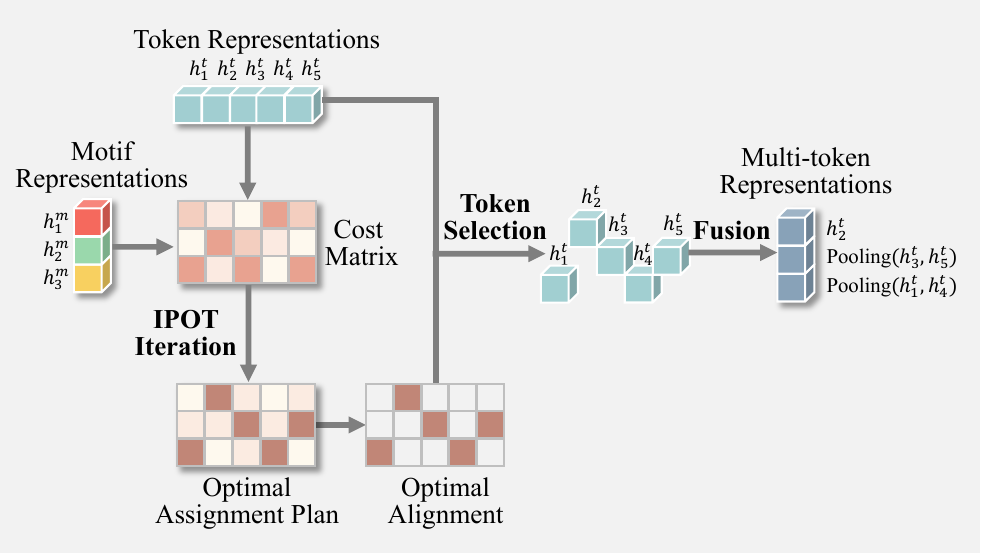}
    \caption{The process of obtaining multi-token representations through OT. Considering the alignments between token representations and motif representations as an optimal transport problem, we fuse the token representations into multi-token representations corresponding to specific motifs.}
    \label{fig:ot}
\end{figure}

As previously mentioned, a crucial aspect of our model is the consideration of the fused representation of multiple tokens aligned with the corresponding motif. 
To this end, we model the alignments between the input tokens and motifs as the Optimal Transport (OT) problem, as illustrated in Figure \ref{fig:ot}.
As a typical machine learning problem, OT aims to find the scheme that minimizes the transportation cost of transferring one distribution to another distribution \cite{peyre2019computational}.
Concretely, we first consider the following crucial definitions:
(1) We regard the token representations $\mathbf H^t$ and motif representations $\mathbf H^m$ as two independent distributions.
(2) $\mathbf C_{ij}=1-\cos(h^t_i, h^m_j)$ is the transportation cost from $h^t_i$ to $h^m_j$, where $\cos(*,*)$ is a cosine distance function.
(3) We define $\mathbf T=\{\mathbf{T}_{ij}\}$, $1$$\leq$$i$$\leq$$N_t$, $1$$\leq$$j$$\leq$$N_m$, as an assignment plan, learned to optimize the alignments between input tokens and motifs.
Solving the optimal transport problem is equivalent to addressing a specific network-flow problem \cite{wasserstein2018luise} as follows:
\begin{equation}
    \begin{aligned}
    \min\sum_{i=1}^{N_t}\sum_{j=1}^{N_m}\mathbf T_{ij}\mathbf C_{ij}=\min\text{Tr}(\mathbf T_{ij}^{\mathsf T}\mathbf C_{ij}),\\
    \end{aligned}
\end{equation}
where $\text{Tr}(\mathbf T_{ij}^{\mathsf T}\mathbf C_{ij})$ represents the Frobenius dot-product.
The exact optimal transport problem generally poses computational challenges due to its complexity. 
To address this challenge, we introduce the Inexact Proximal Point Method for Optimal Transport (IPOT) \cite{ipot-v115-xie20b}, which approximately solves the optimal transport problem to obtain an optimal assignment plan.

Therefore, for the $i$-th token, we determine its optimal alignment motif index as $a(i)=\mathop{\arg\min}\limits_{1\leq j\leq N_m}\text{Tr}(\mathbf T_{ij}^{\mathsf T}\mathbf C_{ij})$.
Then let $D_j$$=$$\{$$i$$|$$a(i)$$=$$j$$,$$1$$\leq$$i$$\leq$$N_t$$\}$ denote the index set, where the corresponding tokens are aligned to the $j$-th motif. 
Subsequently, we average the representations of tokens occurring in $D_j$ to obtain their fused multi-token representation: $h^p_j=\frac{1}{|D_j|}\mathbb{I}(a(i)=j)h^t_i$, where $\mathbb I(\cdot)$ is the indicator function.
Analogously, we can obtain all motif-aligned multi-token representations as
$\mathbf H^p=[h^p_1,h^p_2,\dots,h^p_{N_p}]\in\mathbb R^{N_p\times d}$, where $N_p$ is the number of multi-tokens.

\subsection{Model Training and Inference}
\label{sec:model_training}
Through the above steps, we obtain the representations of input texts and molecules at different granularities. 
Further, we define the following training objective:
\begin{align}
    \mathcal L = \alpha\cdot\mathcal L_{ta} + \beta\cdot\mathcal L_{mm} + (1-\alpha-\beta)\cdot\mathcal L_{sm},
\end{align}
where $\mathcal L_{ta}, \mathcal L_{mm}, \mathcal L_{sm}$ denote the token-atom, multitoken-motif, and sentence-molecule alignment losses, and $\alpha, \beta$ are two coefficients balancing the effects of different losses, respectively.
In the following, we will detail these three losses.

\begin{figure}
    \centering
    \includegraphics[width=0.48\textwidth]{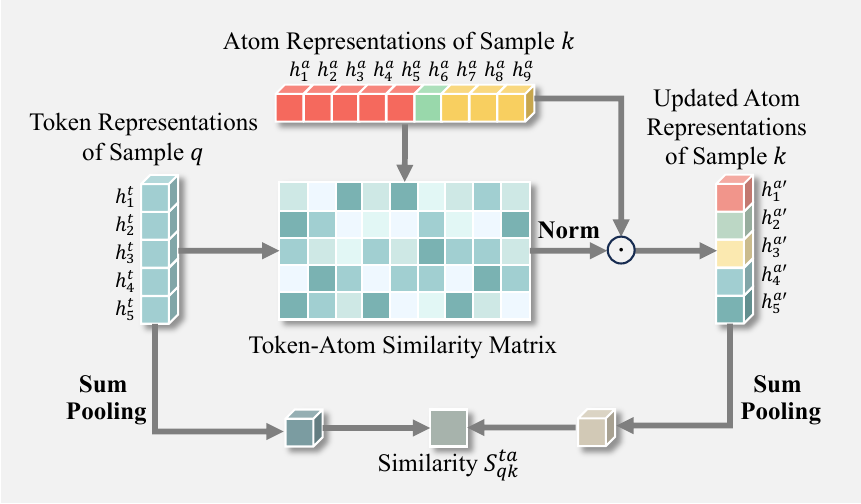}
    \caption{The process of token-atom alignments. Within a batch, we update the atom representations of each sample and then calculate the token-atom level similarities between samples. Then we use contrastive learning to maximize the similarities between matched text-molecule pairs while minimizing those between unmatched pairs.}
    \label{fig:token_atom}
\end{figure}

\subsubsection{Token-Atom Alignment Loss $\mathcal L_{ta}$}
To ensure precise alignments between tokens and atoms, we introduce a token-atom alignment loss based on contrastive learning.
Concretely, as shown in Figure \ref{fig:token_atom}, we compute the token-atom similarity matrix between token representations $\{h^t_i\}$ and atom representations $\{h^a_j\}$.
The similarity matrix is then normalized by the min-max normalization along the atom dimension.
Subsequently, we further normalize the matrix along the token dimension to obtain the alignment weights, used to unify the dimensions of token and atom representations.
After multiplying these alignment weights with atom representations, we update the atom representations as $\{h_i^{a'}\}$ by integrating textual information.
Finally, we apply sum pooling to reduce the dimensions of both representations, which are then used to calculate the final similarity at the token-atom level. 

Within a batch of size $B$, we compute similarities $\{\mathbf{S}_{qk}^{ta}\}$, $1$$\leq$$q$$\leq$$B$, $1$$\leq$$k$$\leq$$B$, between samples at the token-atom level.
Following this, we define the token-atom alignment loss $\mathcal L_{ta}$ as the average of two Categorical Cross-Entropy (CCE) losses corresponding to text-molecule and molecule-text retrieval tasks, respectively:
\begin{align}
    \mathcal L_{ta} = &-\frac{1}{2B}\sum^B_{k=1}\log\frac{\exp(\mathbf{S}^{ta}_{kk})}{\sum^B_{q=1}\exp(\mathbf{S}^{ta}_{qk})}\nonumber\\
    &-\frac{1}{2B}\sum^B_{k=1}\log\frac{\exp(\mathbf{S}^{ta}_{kk})}{\sum^B_{q=1}\exp(\mathbf{S}^{ta}_{kq})}.
\end{align}

\subsubsection{Multitoken-Motif Alignment Loss $\mathcal L_{mm}$}
\label{sec:multitoken-motif}

We also employ contrastive learning to achieve alignments between multi-tokens and motifs.
Specifically, in a similar way,
we calculate the similarities $\{\mathbf S^{mm}_{qk}\}$, $1$$\leq$$q$$\leq$$B$, $1$$\leq$$k$$\leq$$B$, at multi-token and motif level between samples within the same batch.
Thus, the multitoken-motif alignment loss $\mathcal L_{mm}$ is formulated as follows:
\begin{align}
    \mathcal L_{mm}=&-\frac{1}{2B}\sum^B_{k=1}\log\frac{\exp(\mathbf{S}^{mm}_{kk})}{\sum^B_{q=1}\exp(\mathbf{S}^{mm}_{qk})}\nonumber\\
    &-\frac{1}{2B}\sum^B_{k=1}\log\frac{\exp(\mathbf{S}^{mm}_{kk})}{\sum^B_{q=1}\exp(\mathbf{S}^{mm}_{kq})}.
\end{align}

\input{tables/retrieval_chebi}

\subsubsection{Sentence-Molecule Alignment Loss $\mathcal L_{sm}$}
Similar to other levels of alignments, we employ contrastive learning to compute the CCE loss at sentence-molecule level.
Concretely, we maximize the similarities between matched sentence and molecule pairs, while minimizing those between unmatched pairs, with the sentence-molecule alignment loss $\mathcal L_{sm}$ defined as follows:
\begin{align}
    \mathcal L_{sm}=&-\frac{1}{2B}\sum^B_{k=1}\log\frac{\exp(\mathbf{S}^{sm}_{kk})}{\sum^B_{q=1}\exp(\mathbf{S}^{sm}_{qk})}\nonumber\\
    &-\frac{1}{2B}\sum^B_{k=1}\log\frac{\exp(\mathbf{S}^{sm}_{kk})}{\sum^B_{q=1}\exp(\mathbf{S}^{sm}_{kq})},
\end{align}
where $\{\mathbf S^{sm}_{qk}\}$, $1$$\leq$$q$$\leq$$B$, $1$$\leq$$k$$\leq$$B$, is denoted as the similarities between the sentence representation of the $q$-th sample and the molecule representation of the $k$-th sample.

During inference for text-molecule retrieval, we calculate the similarities between text queries and all molecule candidates at the above three levels. Then we assign the coefficients $\alpha, \beta$ to these similarities at different levels to obtain the final similarity, retrieving the molecule with the highest similarity. 
In molecule-text retrieval task, we perform this process in the opposite direction.

%% file: tables/retrieval_chebi.tex
\begin{table*}[!ht]
    \centering
    \caption{Performance on the ChEBI-20 dataset for text-molecule and molecule-text retrieval tasks. The bold part indicates the best performance. ORMA achieves leading performance in both retrieval tasks. $\uparrow$ denotes that the higher is the better, while $\downarrow$ denotes that the lower is the better. $^\dag$ represents our reproduced results of Atomas-base and the other results are reported in the previous works.}
    \begin{tabular}{c|cccc|cccc}
    \toprule
        \multirow{2}{*}{\textbf{Models}} & \multicolumn{4}{c|}{\textbf{Text-Molecule Retrieval}} & \multicolumn{4}{c}{\textbf{Molecule-Text Retrieval}} \\
        \cmidrule(lr){2-5} \cmidrule(lr){6-9} & \textbf{Hits@1($\uparrow$)} & \textbf{Hits@10($\uparrow$)}
        & \textbf{MRR($\uparrow$)} & \textbf{Mean Rank($\downarrow$)} & \textbf{Hits@1($\uparrow$)} & \textbf{Hits@10($\uparrow$)} & \textbf{MRR($\uparrow$)} & \textbf{Mean Rank($\downarrow$)} \\ 
        \midrule
        MLP-Ensemble~\cite{text2mol2021edwards} & 29.4\% & 77.6\% & 0.452 & 20.78 & - & - & - & - \\
        GCN-Ensemble~\cite{text2mol2021edwards} & 29.4\% & 77.1\% & 0.447 & 28.77 & - & - & - & - \\
        All-Ensemble~\cite{text2mol2021edwards} & 34.4\% & 81.1\% & 0.499 & 20.21 & 25.2\% & 74.1\% & 0.408 & 21.77 \\
        \midrule
        MLP+Atten~\cite{text2mol2021edwards} & 22.8\% & 68.7\% & 0.375 & 30.37 & - & - & - & - \\
        MLP+FPG~\cite{fpg2000han} & 22.6\% & 68.6\% & 0.374 & 30.37 & - & - & - & - \\
        \midrule
        AMAN~\cite{adversarial2023zhao} & 49.4\% & 92.1\% & 0.647 & 16.01 & 46.6\% & 91.6\% & 0.625 & 16.50 \\
        Atomas-base$^\dag$~\cite{atomas2024zhang} & 50.1\% & 92.1\% & 0.653 & \textbf{14.49} & 45.6\% & 90.3\% & 0.614 & 15.12 \\
        \midrule
        \rowcolor{cyan!10}\textbf{ORMA (Ours)} & \textbf{66.5\%} & \textbf{93.9\%} & \textbf{0.772} & 18.53 & \textbf{61.6\%} & \textbf{93.8\%} & \textbf{0.739} & \textbf{8.10} \\
    \bottomrule
    \end{tabular}
    \label{tab:retrieval_chebi20}
\end{table*}

%% file: 4-Experiments.tex
\section{Experiments}

\input{tables/retrieval_pcdes}
\input{tables/ablation}

\begin{figure}
    \centering
    \includegraphics[width=0.48\textwidth]{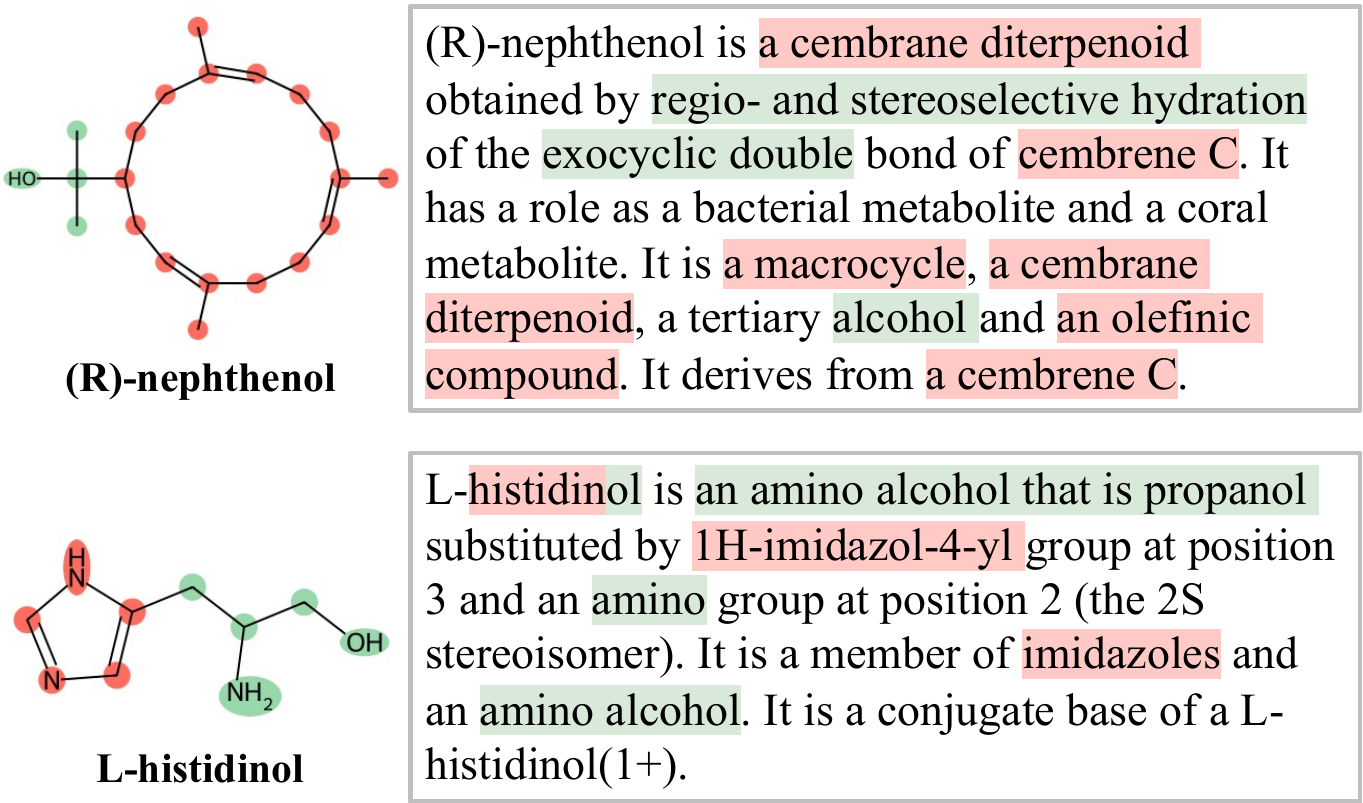}
    \caption{The visualization for the alignments of motifs and multi-tokens. The motifs marked in red and green on the left correspond to the multi-tokens highlighted in red and green on the right, respectively.}
    \label{fig:case_multitoken}
\end{figure}

\subsection{Dataset and Evaluation}
Our model is trained and evaluated on two datasets: ChEBI-20 dataset \cite{text2mol2021edwards} and PCdes dataset \cite{deep2022zeng}. 
The ChEBI-20 dataset contains 33,010 molecules with textual descriptions, divided into training, validation, and test sets in the ratio of 8:1:1. Following previous studies \cite{text2mol2021edwards,adversarial2023zhao}, we evaluate the retrieval performance using Hits@1, Hits@10, mean reciprocal rank (MRR), and mean rank. During inference, test samples are retrieved from the entire ChEBI-20 dataset.
The PCdes dataset consists of 15,000 molecule pairs from PubChem \cite{kim2016pubchem}, split in the ratio of 7:1:2. After excluding 8 molecules whose SMILES strings can not be converted into 2D graphs using RDKit \cite{rdkit}, 14,992 instances remain for our experiments. Following previous works, retrieval performance is evaluated using Recall at 1/10 (R@1, R@10), MRR, and mean rank. Previous pretrained studies \cite{momu2022su, deep2022zeng, molca2023liu, molfm2023luo, moleculestm2023liu, atomas2024zhang} often utilize finetuning or directly inference at zero-shot settings on the PCdes dataset. Since our approach does not involve pretraining, we train our model from scratch on the PCdes training set and evaluate its performance on the test set.

\subsection{Experimental Settings}
During model training, we utilize a pretrained SciBERT\cite{scibert2019beltagy} as the text encoder with a maximum text length of 256. We employ a 3-layer GCN as the graph encoder with an output dimension of 300. We use the Adam optimizer \cite{kingma2014adam}, setting the learning rate of 3e-5 for SciBERT and 1e-4 for the remaining parts of our model. Our training spans 60 epochs and uses a batch size of 32. We assign the coefficients for different levels in training objective as follows: $\alpha = 0.5, \beta = 0.2$.

\subsection{Baseline Models}

\paragraph{AMAN \cite{adversarial2023zhao}}
It leverages the SciBERT to encode textual descriptions and a Graph Transformer Network (GTN) to encode molecules. Besides, it introduces adversarial learning and triplet loss to align these modalities.

\paragraph{Atomas \cite{atomas2024zhang}}
This model is pretrained on large-scale text and SMILES data, aligning textual and molecular modalities at three granularities.
Atomas utilizes the clustering approach to align textual descriptions and molecules at three granularities, while our model adopts optimal transport and contrastive learning to achieve alignments at three levels.
It inferences at zero-shot settings for evaluating retrieval performance on PCdes. Additionally, on the ChEBI-20 dataset, we train Atomas from scratch using the same configuration as ours to evaluate its retrieval capabilities.

\begin{figure*} 
    \centering
    \includegraphics[width=0.9\textwidth]{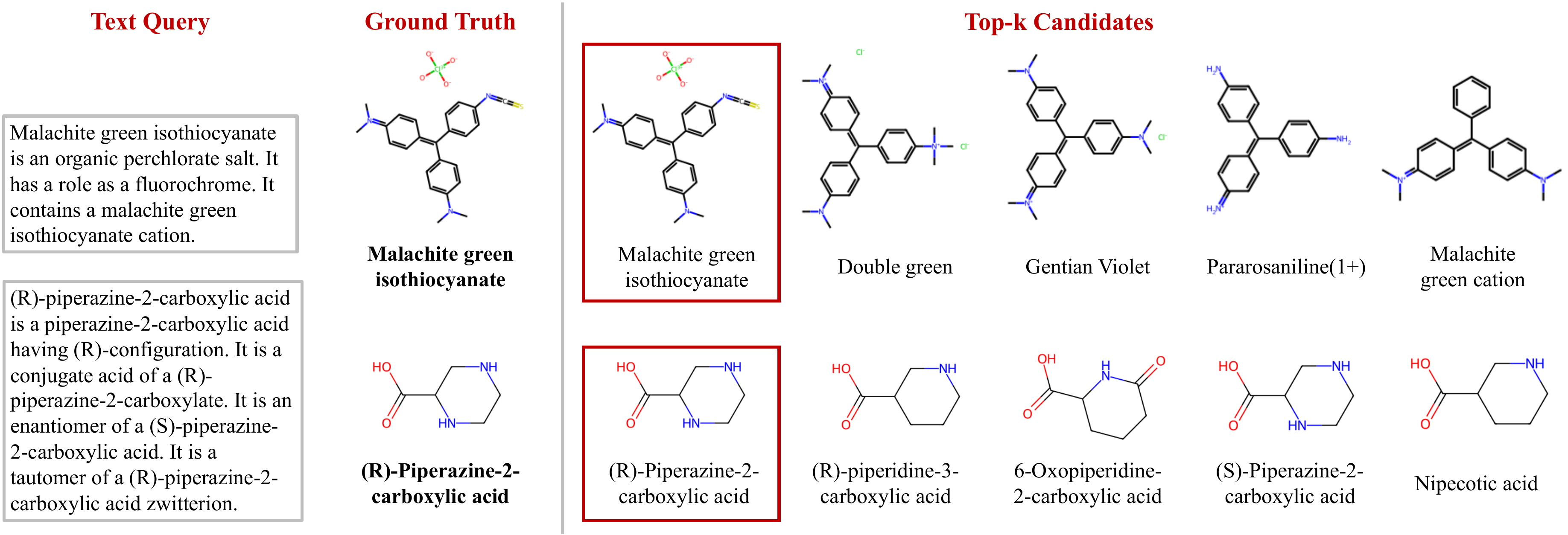}
    \caption{Case study of our model in the text-molecule retrieval task. 
    The red box indicates that the molecule with the highest similarity retrieved by our model is the ground truth.}
    \label{fig:case_text2mol}
\end{figure*}

\begin{figure*}
    \centering
    \includegraphics[width=0.9\textwidth]{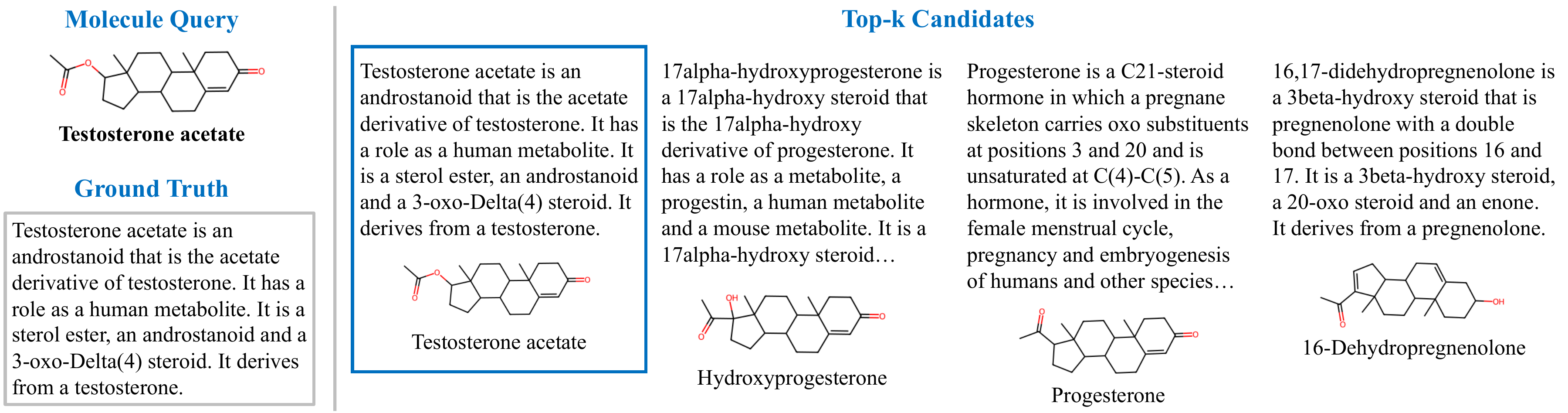}
    \caption{Case study of our model in the molecule-text retrieval task. 
    The blue box highlights the candidate description retrieved by our model that exactly corresponds to the ground truth.}
    \label{fig:case_mol2text}
\end{figure*}

\subsection{Main Results}

We present experimental results on the ChEBI-20 dataset for both retrieval tasks, as detailed in Table \ref{tab:retrieval_chebi20}. 
In the text-molecule retrieval task, our model significantly outperforms all other baseline models, achieving a Hits@1 of 66.5\%, Hits@10 of 93.9\%, and an MRR of 0.772. 
Although our mean rank is 18.53 that is not the best, it is competitively close to the best-performing model, Atomas-base, which obtains a slightly better mean rank of 14.49 but lower Hits@1 and MRR.

In the molecule-text retrieval task, our model also demonstrates superior performance, leading in all metrics with a Hits@1 of 61.6\%, Hits@10 of 93.8\%, an MRR of 0.739, and an impressively low mean rank of 8.10. 
The experimental results confirm the effectiveness of our model in these retrieval tasks, providing valuable insights for further enhancements.

We also present experimental results on the PCdes dataset for two retrieval tasks, as detailed in Table \ref{tab:retrieval_pcdes}. 
Notably, our model, trained from scratch on this dataset, significantly outperforms existing models at both zero-shot and finetuning settings. 
In the text-molecule retrieval task, our model achieves an impressive R@1 of 64.8\%, remarkably better than the best zero-shot model (Atomas-large at 49.1\%) and finetuning model (MolFM at 29.8\%). It also achieves an R@5 of 82.3\%, R@10 of 86.3\%, and MRR of 0.727. 
In the molecule-text retrieval task, our model demonstrates superior results with an R@1 of 62.1\%, R@5 of 81.4\%, and R@10 of 86.3\%, alongside MRR of 0.710. 
This exceptional performance highlights the potential of our model to surpass generalized pretrained models in both retrieval tasks, confirming its effectiveness.

\subsection{Ablation Study}
We conduct an ablation study on the ChEBI-20 dataset for both retrieval tasks, providing insights about the benefits of alignments at different levels: token-atom (TA), multitoken-motif (MM), and sentence-molecule (SM).
From Table \ref{tab:ablation}, we can clearly observe that the use of all three levels leads to the best results, confirming the effectiveness of the multi-level alignments. Conversely, the results are worse when only two levels of alignments are employed. 
Also, relying on one level of alignment alone further diminishes the performance.
These results highlight that alignment at each level uniquely contributes to the overall performance.

\subsection{Case Study}
\subsubsection{Retrieval Performance}
We select several typical samples from the test set to demonstrate the retrieval capability of our model. 
As illustrated in Figure \ref{fig:case_text2mol}, our model accurately retrieves the ground truth molecules as the top-1 candidate for both queries. The molecules retrieved in the top-k candidates closely resemble the text queries, confirming the effective alignments between textual descriptions and molecules.
As shown in Figure \ref{fig:case_mol2text}, our model also correctly identifies the ground truth description as the top-1 retrieval result for a given molecule query. Although the textual descriptions of the top-k retrieved candidates vary, their corresponding molecule structures are remarkably similar. This suggests that our model accurately grasps the alignments between text and molecules, highlighting its superior performance.

\subsubsection{Visualization for the Alignments of Motifs and Multi-tokens}
We also visualize the fusion of multiple motif-aligned tokens through optimal transport theory in Figure \ref{fig:case_multitoken}. On the left side of the figure, atoms marked in the same color serve as the motifs extracted by our model.
On the right side, the textual descriptions highlighted in the same colors represent the identified multi-tokens corresponding to the motifs of matching colors on the left. 
For instance, we use \textit{(R)-nephthenol} and \textit{L-histidinol} as examples, where the motifs extracted by our model match the identified multi-tokens.
This comprehensively demonstrates that the optimal transport is helpful for the alignments between the multi-tokens and corresponding motifs. 

%% file: tables/retrieval_pcdes.tex
\begin{table*}[!ht]
    \centering
    \caption{Results on the PCdes dataset for text-molecule and molecule-text retrieval tasks at zero-shot, finetuning and from-scratch settings. The bold parts indicate the best performance. Note that our model, trained from scratch, significantly outperforms others in all metrics. $\uparrow$ denotes that the higher is the better.}
    \begin{tabular}{c|cccc|cccc}
    \toprule
        \multirow{2}{*}{\textbf{Models}} & \multicolumn{4}{c|}{\textbf{Text-Molecule Retrieval}} & \multicolumn{4}{c}{\textbf{Molecule-Text Retrieval}} \\
        \cmidrule(lr){2-5} \cmidrule(lr){6-9} & \textbf{R@1($\uparrow$)} & \textbf{R@5($\uparrow$)} & \textbf{R@10($\uparrow$)} & \textbf{MRR($\uparrow$)} & \textbf{R@1($\uparrow$)} & \textbf{R@5($\uparrow$)} & \textbf{R@10($\uparrow$)} & \textbf{MRR($\uparrow$)} \\ 
        \midrule
        \rowcolor{gray!20}\multicolumn{9}{l}{\textit{Pretrained Model + Zero-shot}} \\
        MoMu~\cite{momu2022su} & 4.9\% & 14.5\% & 20.7\% & 0.103 & 5.1\% & 12.8\% & 18.9\% & 0.099 \\
        MolCA~\cite{molca2023liu} & 35.1\% & 62.1\% & 69.8\% & 0.473 & 38.0\% & 66.8\% & 74.5\% & 0.508 \\
        MolFM~\cite{molfm2023luo} & 16.1\% & 30.7\% & 39.5\% & 0.236 & 13.9\% & 28.7\% & 36.2\% & 0.214 \\
        MoleculeSTM~\cite{moleculestm2023liu} & 35.8\% & - & - & - & 39.5\% & - & - & - \\
        Atomas-base~\cite{atomas2024zhang} & 39.1\% & 59.7\% & 66.6\% & 0.473 & 37.9\% & 59.2\% & 65.6\% & 0.478 \\
        Atomas-large~\cite{atomas2024zhang} & 49.1\% & 68.3\% & 73.2\% & 0.578 & 46.2\% & 66.0\% & 72.3\% & 0.555 \\
        \midrule
        \rowcolor{gray!20}\multicolumn{9}{l}{\textit{Pretrained Model + Finetuning}} \\
        SciBERT~\cite{scibert2019beltagy} & 16.3\% & 33.9\% & 42.6\% & 0.250 & 15.0\% & 34.1\% & 41.7\% & 0.239 \\
        KV-PLM~\cite{deep2022zeng} & 18.4\% & 37.2\% & 45.4\% & 0.274 & 16.6\% & 35.9\% & 44.8\% & 0.260 \\
        KV-PLM*~\cite{deep2022zeng} & 20.6\% & 37.9\% & 45.7\% & 0.292 & 19.3\% & 37.3\% & 45.3\% & 0.281 \\
        MoMu~\cite{momu2022su} & 24.5\% & 45.4\% & 53.8\% & 0.343 & 24.9\% & 44.9\% & 54.3\% & 0.345 \\
        MolFM~\cite{molfm2023luo} & 29.8\% & 50.5\% & 58.6\% & 0.396 & 29.4\% & 50.3\% & 58.5\% & 0.393 \\
        \midrule
        \rowcolor{gray!20}\multicolumn{9}{l}{\textit{From-scratch}} \\
        \rowcolor{cyan!10}\textbf{ORMA (Ours)} & \textbf{64.8\%} & \textbf{82.3\%} & \textbf{86.3\%} & \textbf{0.727} & \textbf{62.1\%} & \textbf{81.4\%} & \textbf{86.3\%} & \textbf{0.710} \\
    \bottomrule
    \end{tabular}
    \label{tab:retrieval_pcdes}
\end{table*}

%% file: tables/ablation.tex
\begin{table*}[ht]
    \centering
    \caption{Ablation study on the ChEBI-20 dataset at three levels: token-atom (TA), Multitoken-motif (MM), and sentence-molecule (SM). The results suggest that integrating three levels significantly improves performance in both text-molecule and molecule-text retrieval tasks. $\downarrow$ denotes that the lower is the better.
    }
    \resizebox{0.9\textwidth}{!}{
    \begin{tabular}{ccc|cccc|cccc}
        \toprule
        \multicolumn{3}{c|}{\textbf{Level}} & \multicolumn{4}{c|}{\textbf{Text-Molecule Retrieval}} & \multicolumn{4}{c}{\textbf{Molecule-Text Retrieval}} \\
        \cmidrule(lr){1-3} \cmidrule(lr){4-7} \cmidrule(lr){8-11}
        \textbf{TA} & \textbf{MM} & \textbf{SM} & \textbf{Hits@1($\uparrow$)} & \textbf{Hits@10($\uparrow$)} & \textbf{MRR($\uparrow$)} & \textbf{Mean Rank($\downarrow$)} & \textbf{Hits@1($\uparrow$)} & \textbf{Hits@10($\uparrow$)} & \textbf{MRR($\uparrow$)} & \textbf{Mean Rank($\downarrow$)} \\
        \midrule
        \cmark & \cmark & \cmark & \textbf{66.5\%} & \textbf{93.9\%} & \textbf{0.772} & \textbf{18.53} & \textbf{61.6\%} & \textbf{93.8\%} & \textbf{0.739} & \textbf{8.1} \\
        \midrule
        \cmark & \cmark &  & 57.4\% & 91.6\% & 0.700 & 26.74 & 38.4\% & 89.4\% & 0.552 & 12.87 \\
        \cmark &  & \cmark & 61.7\% & 92.5\% & 0.733 & 23.10 & 51.9\% & 91.3\% & 0.661 & 11.13 \\
         & \cmark & \cmark & 35.6\% & 81.8\% & 0.511 & 41.39 & 35.1\% & 80.8\% & 0.506 & 30.95 \\
        \midrule
         &  & \cmark & 24.1\% & 75.3\% & 0.401 & 42.28 & 21.8\% & 72.1\% & 0.377 & 30.86 \\
        & \cmark &  & 17.0\% & 55.0\% & 0.291 & 219.45 & 20.2\% & 56.3\% & 0.319 & 247.27 \\
        \cmark &  &  & 49.6\% & 89.1\% & 0.636 & 37.17 & 23.3\% & 71.6\% & 0.384 & 22.64 \\
        \bottomrule
    \end{tabular}
    }
    \label{tab:ablation}
\end{table*}

%% file: 5-Conclusion.tex
\section{Conclusion}

In this paper, we propose an \textbf{O}ptimal T\textbf{R}ansport-based \textbf{M}ulti-grained \textbf{A}lignments model (\textbf{ORMA}), designed to enhance the cross-modal text-molecule retrieval task. 
By utilizing SciBERT, we encode textual descriptions into token and sentence representations.
Simultaneously, we represent the molecule as a heterogeneous graph containing atom, motif, and molecule nodes, encoded by our GCN molecule encoder.
Innovatively, we consider the alignments between tokens and motifs as the optimal transport problem, fusing token representations into motif-aligned multi-token representations.
Furthermore, we employ contrastive learning to align textual descriptions and molecules at three levels: token-atom, multitoken-motif, and sentence-molecule. 
To the best of our knowledge, our work is the first attempt to consider representations at the motif and multi-token granularities.
Experimental results on the ChEBI-20 and PCdes datasets demonstrate the superior retrieval performance of our model compared to previous state-of-the-art models. 
In the future, we will extend ORMA to integrate more modal data sources, such as protein structures and cellular images, to further the application of ORMA in more complex biological systems.

%% file: main.bbl
\begin{thebibliography}{10}

\bibitem{kim2016pubchem}
S.~Kim, P.~A. Thiessen, E.~E. Bolton, J.~Chen, G.~Fu, A.~Gindulyte, L.~Han, J.~He, S.~He, B.~A. Shoemaker, {\em et~al.}, ``Pubchem substance and compound databases,'' {\em Nucleic acids research}, vol.~44, no.~D1, pp.~D1202--D1213, 2016.

\bibitem{text2mol2021edwards}
C.~Edwards, C.~Zhai, and H.~Ji, ``Text2mol: Cross-modal molecule retrieval with natural language queries,'' in {\em Proceedings of the 2021 Conference on Empirical Methods in Natural Language Processing}, pp.~595--607, 2021.

\bibitem{deep2022zeng}
Z.~Zeng, Y.~Yao, Z.~Liu, and M.~Sun, ``A deep-learning system bridging molecule structure and biomedical text with comprehension comparable to human professionals,'' {\em Nature communications}, vol.~13, no.~1, p.~862, 2022.

\bibitem{molt52022edwards}
C.~Edwards, T.~Lai, K.~Ros, G.~Honke, K.~Cho, and H.~Ji, ``Translation between molecules and natural language,'' {\em arXiv preprint arXiv:2204.11817}, 2022.

\bibitem{molxpt2023liu}
Z.~Liu, W.~Zhang, Y.~Xia, L.~Wu, S.~Xie, T.~Qin, M.~Zhang, and T.-Y. Liu, ``Molxpt: Wrapping molecules with text for generative pre-training,'' {\em arXiv preprint arXiv:2305.10688}, 2023.

\bibitem{smiles1998weininger}
D.~Weininger, ``Smiles, a chemical language and information system. 1. introduction to methodology and encoding rules,'' {\em Journal of chemical information and computer sciences}, vol.~28, no.~1, pp.~31--36, 1988.

\bibitem{momu2022su}
B.~Su, D.~Du, Z.~Yang, Y.~Zhou, J.~Li, A.~Rao, H.~Sun, Z.~Lu, and J.-R. Wen, ``A molecular multimodal foundation model associating molecule graphs with natural language,'' {\em arXiv preprint arXiv:2209.05481}, 2022.

\bibitem{moleculestm2023liu}
S.~Liu, W.~Nie, C.~Wang, J.~Lu, Z.~Qiao, L.~Liu, J.~Tang, C.~Xiao, and A.~Anandkumar, ``Multi-modal molecule structure--text model for text-based retrieval and editing,'' {\em Nature Machine Intelligence}, vol.~5, no.~12, pp.~1447--1457, 2023.

\bibitem{adversarial2023zhao}
W.~Zhao, D.~Zhou, B.~Cao, K.~Zhang, and J.~Chen, ``Adversarial modality alignment network for cross-modal molecule retrieval,'' {\em IEEE Transactions on Artificial Intelligence}, 2023.

\bibitem{atomas2024zhang}
Y.~Zhang, G.~Ye, C.~Yuan, B.~Han, L.-K. Huang, J.~Yao, W.~Liu, and Y.~Rong, ``Atomas: Hierarchical alignment on molecule-text for unified molecule understanding and generation,'' {\em arXiv preprint arXiv:2404.16880}, 2024.

\bibitem{unifying2023christofidellis}
D.~Christofidellis, G.~Giannone, J.~Born, O.~Winther, T.~Laino, and M.~Manica, ``Unifying molecular and textual representations via multi-task language modelling,'' in {\em International Conference on Machine Learning}, pp.~6140--6157, PMLR, 2023.

\bibitem{molca2023liu}
Z.~Liu, S.~Li, Y.~Luo, H.~Fei, Y.~Cao, K.~Kawaguchi, X.~Wang, and T.-S. Chua, ``Molca: Molecular graph-language modeling with cross-modal projector and uni-modal adapter,'' {\em arXiv preprint arXiv:2310.12798}, 2023.

\bibitem{2024arXiv241023715S}
J.~{Song}, W.~{Zhuang}, Y.~{Lin}, L.~{Zhang}, C.~{Li}, J.~{Su}, S.~{He}, and X.~{Bo}, ``{Towards Cross-Modal Text-Molecule Retrieval with Better Modality Alignment},'' {\em arXiv e-prints}, p.~arXiv:2410.23715, Oct. 2024.

\bibitem{molfm2023luo}
Y.~Luo, K.~Yang, M.~Hong, X.~Liu, and Z.~Nie, ``Molfm: A multimodal molecular foundation model,'' {\em arXiv preprint arXiv:2307.09484}, 2023.

\bibitem{git2024liu}
P.~Liu, Y.~Ren, J.~Tao, and Z.~Ren, ``Git-mol: A multi-modal large language model for molecular science with graph, image, and text,'' {\em Computers in Biology and Medicine}, vol.~171, p.~108073, 2024.

\bibitem{3dmollm2024li}
S.~Li, Z.~Liu, Y.~Luo, X.~Wang, X.~He, K.~Kawaguchi, T.-S. Chua, and Q.~Tian, ``Towards 3d molecule-text interpretation in language models,'' {\em arXiv preprint arXiv:2401.13923}, 2024.

\bibitem{fan2018multi}
F.~Fan, Y.~Feng, and D.~Zhao, ``Multi-grained attention network for aspect-level sentiment classification,'' in {\em Proceedings of the 2018 conference on empirical methods in natural language processing}, pp.~3433--3442, 2018.

\bibitem{zhang2018alignment}
B.~Zhang, D.~Xiong, J.~Su, and Y.~Qin, ``Alignment-supervised bidimensional attention-based recursive autoencoders for bilingual phrase representation,'' {\em IEEE transactions on cybernetics}, vol.~50, no.~2, pp.~503--513, 2018.

\bibitem{zhang2017battrae}
B.~Zhang, D.~Xiong, and J.~Su, ``Battrae: Bidimensional attention-based recursive autoencoders for learning bilingual phrase embeddings,'' in {\em Proceedings of the AAAI conference on artificial intelligence}, vol.~31, 2017.

\bibitem{su2015bilingual}
J.~Su, D.~Xiong, B.~Zhang, Y.~Liu, J.~Yao, and M.~Zhang, ``Bilingual correspondence recursive autoencoder for statistical machine translation,'' in {\em Proceedings of the 2015 Conference on Empirical Methods in Natural Language Processing}, pp.~1248--1258, 2015.

\bibitem{su2016convolution}
J.~Su, B.~Zhang, D.~Xiong, R.~Li, and J.~Yin, ``Convolution-enhanced bilingual recursive neural network for bilingual semantic modeling,'' in {\em Proceedings of COLING 2016, the 26th International Conference on Computational Linguistics: Technical Papers}, pp.~3071--3081, 2016.

\bibitem{zhang2016bilingual}
B.~Zhang, D.~Xiong, J.~Su, H.~Duan, and M.~Zhang, ``Bilingual autoencoders with global descriptors for modeling parallel sentences,'' in {\em Proceedings of COLING 2016, the 26th International Conference on Computational Linguistics: Technical Papers}, pp.~2548--2558, 2016.

\bibitem{xclip2022ma}
Y.~Ma, G.~Xu, X.~Sun, M.~Yan, J.~Zhang, and R.~Ji, ``X-clip: End-to-end multi-grained contrastive learning for video-text retrieval,'' in {\em Proceedings of the 30th ACM International Conference on Multimedia}, pp.~638--647, 2022.

\bibitem{jin2023dicosa}
P.~Jin, H.~Li, Z.~Cheng, J.~Huang, Z.~Wang, L.~Yuan, C.~Liu, and J.~Chen, ``Text-video retrieval with disentangled conceptualization and set-to-set alignment,'' {\em arXiv preprint arXiv:2305.12218}, 2023.

\bibitem{wang2023ucofia}
Z.~Wang, Y.-L. Sung, F.~Cheng, G.~Bertasius, and M.~Bansal, ``Unified coarse-to-fine alignment for video-text retrieval,'' in {\em Proceedings of the IEEE/CVF International Conference on Computer Vision}, pp.~2816--2827, 2023.

\bibitem{scibert2019beltagy}
I.~Beltagy, K.~Lo, and A.~Cohan, ``Scibert: A pretrained language model for scientific text,'' {\em arXiv preprint arXiv:1903.10676}, 2019.

\bibitem{hierarchical2023zang}
X.~Zang, X.~Zhao, and B.~Tang, ``Hierarchical molecular graph self-supervised learning for property prediction,'' {\em Communications Chemistry}, vol.~6, no.~1, p.~34, 2023.

\bibitem{brics2008degen}
J.~Degen, C.~Wegscheid-Gerlach, A.~Zaliani, and M.~Rarey, ``On the art of compiling and using'drug-like'chemical fragment spaces,'' {\em ChemMedChem}, vol.~3, no.~10, p.~1503, 2008.

\bibitem{gcn2016kipf}
T.~N. Kipf and M.~Welling, ``Semi-supervised classification with graph convolutional networks,'' {\em arXiv preprint arXiv:1609.02907}, 2016.

\bibitem{peyre2019computational}
G.~Peyr{\'e}, M.~Cuturi, {\em et~al.}, ``Computational optimal transport: With applications to data science,'' {\em Foundations and Trends{\textregistered} in Machine Learning}, vol.~11, no.~5-6, pp.~355--607, 2019.

\bibitem{wasserstein2018luise}
G.~Luise, A.~Rudi, M.~Pontil, and C.~Ciliberto, ``Differential properties of sinkhorn approximation for learning with wasserstein distance,'' {\em Advances in Neural Information Processing Systems}, vol.~31, 2018.

\bibitem{ipot-v115-xie20b}
Y.~Xie, X.~Wang, R.~Wang, and H.~Zha, ``A fast proximal point method for computing exact wasserstein distance,'' in {\em Proceedings of The 35th Uncertainty in Artificial Intelligence Conference} (R.~P. Adams and V.~Gogate, eds.), vol.~115 of {\em Proceedings of Machine Learning Research}, pp.~433--453, PMLR, 22--25 Jul 2020.

\bibitem{fpg2000han}
J.~Han and J.~Pei, ``Mining frequent patterns by pattern-growth: methodology and implications,'' {\em ACM SIGKDD explorations newsletter}, vol.~2, no.~2, pp.~14--20, 2000.

\bibitem{rdkit}
{RDKit}, ``Rdkit: Open-source cheminformatics.''
\newblock \url{https://www.rdkit.org}.

\bibitem{kingma2014adam}
D.~P. Kingma and J.~Ba, ``Adam: A method for stochastic optimization,'' {\em arXiv preprint arXiv:1412.6980}, 2014.

\end{thebibliography}
